\documentclass[11pt]{article}
\pdfoutput=1
\usepackage{jcappub,natbib,float,verbatim}
\usepackage{color}
\usepackage{tikz,booktabs,makecell}
\usepackage{array,graphicx,subfig}

\usepackage{jcappub} 

\def\mnras{Monthly Notices of the Royal Astronomical Society}
\def\apj{The Astrophysical Journal}

\def\jcap{Journal of Cosmology and Astroparticle Physics}
\def\prd{Physical Review D}
\def\aap{Astronomy $\&$ Astrophysics}

\def\nat{Nature}

\title{Cosmological inference from standard sirens without redshift measurements}
\author{Xuheng Ding$^{a,b,c}$, Marek Biesiada$^{c,d}$, Xiaogang Zheng$^{c,d}$, Kai Liao$^{e}$, Zhengxiang Li$^{c}$, Zong-Hong Zhu$^{a,c}$}
\affiliation[a]{School of Physics and Technology, Wuhan University, Wuhan 430072, China}
\affiliation[b]{Department of Physics and Astronomy, University of California, Los Angeles, CA, 90095-1547, USA}
\affiliation[c]{Department of Astronomy, Beijing Normal University, Beijing 100875, China}
\affiliation[d]{Department of Astrophysics and Cosmology, Institute of Physics, University of Silesia, 75 Pu{\l}ku Piechoty 1, 41-500 Chorz{\'o}w,
Poland}
\affiliation[e]{School of Science, Wuhan University of Technology, Wuhan 430070, China}
\emailAdd{dingxh@whu.edu.cn}

\abstract{
The purpose of this work is to investigate the prospects of using the future standard siren
data without redshift measurements to constrain cosmological parameters.
With successful detections of gravitational wave (GW) signals an era of GW astronomy has
begun. Unlike the electromagnetic domain, GW signals allow direct measurements of
luminosity distances to the sources, while their redshifts remain to be measured by
identifying electromagnetic counterparts. This leads to significant technical problems for almost all
possible BH-BH systems. It is the major obstacle to cosmological applications of GW standard
sirens.
In this paper, we introduce the general framework of using luminosity distances alone for
cosmological inference. The idea is to use the prior knowledge of the
redshift probability distribution for coalescing
sources from the intrinsic merger rates assessed with population synthesis codes.
Then the posterior probability distributions for cosmological parameters can be calculated.
We demonstrate the performance of our method on the simulated mock data and
show that the luminosity distance measurement would enable an
accurate determination of cosmological parameters up to $20\%$ uncertainty level.
We also find that in order to infer $H_0$ to 1\% level with flat $\Lambda$CDM model, we
need about $10^5$ events.}


\begin{document}
\maketitle
\flushbottom
\section{Introduction}
\label{into}
In the past decade, flat $\Lambda$CDM model has emerged as the standard cosmological model. By assuming the existence of some form of dark energy with equation of state coefficient $w = -1$ (equivalent to the cosmological constant) and the cold dark matter (CDM) predominantly clumped in large scale structure, the flat $\Lambda$CDM model is consistent with various cosmological observations. Two key parameters in this model are the Hubble constant $H_0$ and the matter density parameter $\Omega_m$. The $H_0$ is particularly noteworthy which represents the current expansion rate and is related to the age, size, and critical density of the Universe. However, the inferred value of $H_0$ from the \textit{Planck} satellite \cite{Planck2015} is in tension with other low redshift measurements \cite{Riess2016, Freedman2012, Bonvin2016}. Thus, a 1\% accurate measurement of $H_0$ is highly needed to understand whether the tensions within the $\Lambda$CDM model are real and require new physics.

Recent detections of gravitational waves (GW) by advanced LIGO/Virgo detectors have opened a new window to the Universe \cite{Abbott1, Abbott2, Abbott3,GW170814,GW170817,GW170608} which will have a significant impact on physics and astronomy. The detected GW signal came from the coalescence of double compact objects (DCO)\footnote{The DCO comprise of NS-NS, BH-NS and BH-BH binary systems.}, i.e., BH-BH binary systems in all these cases, besides GW170817 \citep{GW170817} which was the first  NS-NS coalescence ever detected and accompanied by successfully identified electromagnetic (EM) counterpart.
This category of GW sources (i.e., inspiralling DCOs) can be considered as  {\it standard sirens} \cite{Schutz1986} ---  named so,  in analogy to {\it standard candles} in the EM domain.
However, contrary to EM probes, with GW inspiral signal one can measure the luminosity distance $D_L$ to the source directly, without the need of taking into account the cosmic distance ladder.
This idea has been widely discussed in the literature for both ground-based detectors \cite{Nissanke2010, Zhao2011, Pozzo2012, Taylor12, Cai2017} and space-based detectors \cite{Holz2005, Tamanini2016} --- the above references being just indicative and by no means exhaustive.

Encouraged by the development of new technologies leading to the Advanced LIGO detectors, GW scientific community is designing and planning to build a new generation detector called the Einstein Telescope (ET) which will broaden the accessible volume of the Universe by three orders of magnitude. Given this sensitivity, the service of ET would yield $10^3 -10^7$ detections per year up to redshift $z=17$ \cite{Abernathy2011}, and thus could provide a considerable database of luminosity distances to these sources. Such rich statistics of $D_L$ measurements is very promising in the context of constraining cosmological parameters to much higher precision. In particular, it was demonstrated that with a dozen of lensed GW and EM signals, which would be quite realistic in the era of the ET, one could measure $H_0$ with
sub-percent accuracy \cite{kai2017}.

The standard way of cosmological inference is by using the $D_L-z$ diagram and confronting observed values with theoretical $D_L(z)$ function dependent on cosmological parameters.  Therefore one needs an independent determination of $D_L$ and the redshift $z$. In the EM domain redshift is an obvious observable, with $D_L$ being a tricky one. With GW signals, the reverse is true: $D_L$ is a direct observable, while $z$ should somehow be assessed. In a vast majority of works discussing the standard siren approach, an optimistic assumption was made that an accompanying EM signal will be detected thus allowing for determination of $z$. However it is not an easy task, first because the GW signals are detected with low resolution of their sky location, typically $\sim 10$ deg$^2$ accuracy  \cite{Abbott+2017}. Hence, identifying the EM counterpart from the extensive region is difficult. Fortunately, in the case of GW170817 favorable location of the source with respect to three LIGO/Virgo detectors and the luminosity distance inferred from the waveform considerably constrained candidate host galaxies and eventually optical counterpart was observed in one of them.  In several papers, it was proposed to determine the  $\Delta z_i$ range from the possible host galaxies for the GW source \cite{MacLeod2008, Pozzo2012, Pozzo17}.
Similarly, the cross-correlation between spatial distributions of the DCOs and the known-redshift
galaxies was proposed to constrain the distance-redshift relation \cite{Oguri2016}. The authors of \cite{Arabsalmani2013} suggested a self-calibrating iterative scheme to mitigate the misidentification of DCO sources.
However, some of the GW signals registered in the era of the ET would come from very high redshifts (i.e., $z>4$) where the EM counterparts and possible hosts are extremely faint and not readily observable. At last, the DCO signals are dominated by the BH-BH systems which most likely would not be accompanied by noticeable EM counterparts. This is what could be expected on theoretical grounds and what we indeed experienced with five successful BH-BH detections so far.

Attempting to overcome the difficulties with redshifts inherent to GW astrophysics, we propose a new approach to construct the posterior probability distribution for the cosmological parameters using the distribution of sources' redshifts as a prior.
In this paper, we calculate the prior redshift distribution of sources based on the intrinsic merger rates of the DCOs together with the expected sensitivity of the ET.
On the simulated data, we show that one could achieve the precision of cosmological inference comparable to that achievable from current EM data and one would be able to  measure $H_0$ with 1\%
accuracy using the data gathered by the ET in one year of its operation.

The paper is organized as follows. In Section \ref{sec:theory}, we outline the method of constructing the posterior  probability distribution for cosmological parameters and introduce an idea of how to set up the priors. In Section \ref{sim&res}, we investigate the prospects of our approach by carrying out the Monte Carlo probability maximization using simulated mock data for two popular cosmological models.
We discuss the results and conclude in Section \ref{sec:conclusion}.

\section{Methodology}
\label{sec:theory}
\subsection{Theoretical framework}
In this section, we introduce the general framework of using luminosity distances alone for cosmological inference. Usually, the redshift is a key information in this context, but as we already mentioned for most of GW events redshift will be unknown and unmeasurable. Therefore, we start with noticing that the redshift probability distribution for DCO coalescing sources could be calculated and then used to derive posterior probability distributions for cosmological parameters.

Let us consider a number of $n$ GW events detected by the ET, 
with luminosity distances directly measured from their waveforms denoted collectively as $ \vec{D}\equiv (D_{1}, D_{2}, ..., D_{n} )$. The redshifts of these events are unknown. Our goal is to construct the posterior probability distribution of cosmological parameters $\vec{\Omega} \equiv (H_0, \Omega_m$) for $\Lambda$CDM model (or some bigger collection of parameters for other cosmological models). We will use Bayes theorem taking the redshift distribution of the sources as a prior.

Focusing on the $i$-th event 
one can write:
\begin{eqnarray}
\label{Baye2}
  P(\vec{\Omega},z_i|D_{i},I) &~=~& P(\vec{\Omega}|z_i,D_{i},I) P_{obs}(z_i|D_i,I)  \nonumber\\
&~=~& \frac{P(D_{i}|\vec{\Omega},z_i,I)P(\vec{\Omega}|z_i,I)}{P(D_i|z_i,I)}
\frac{P(D_i|z_i,I)P_{obs}(z_i|I)}{P(D_i|I)}  \nonumber\\
&~=~&\frac{P(D_i|\vec{\Omega},z_i,I)P_{obs}(z_i|\vec{\Omega},I)}{P(D_i|I)}P(\vec{\Omega}|I)
\end{eqnarray}
where $P(D_i|\vec{\Omega},z_i,I)$ is the likelihood function for the observed data. $P_{obs}(z_i|\vec{\Omega},I)$ and $P(\vec{\Omega}|I)$ are the priors on the redshift and the cosmological parameters, respectively.
Let us emphasize that the prior on redshifts is a prior of observed events and already includes detector selection effects. Hence we used the notation
$P_{obs}(z_i|...)$ to make it clear to the reader that it does not represent intrinsic redshift distribution of sources.
All the other background information related to this study is denoted by $I$ and all probabilities considered are conditional on it. As usual, $P(D_i|I)$ plays the role of normalization constant.
Note that the expression of $P_{obs}(z_i|\vec{\Omega},I)$ means that the redshift probability distribution could be inferred invoking a specific cosmological model.
The likelihood $P(D_i|\vec{\Omega},z_i,I)$ can be taken in the form:
\begin{equation}
\label{likeli}
P(D_i|\vec{\Omega},z_i,I) \propto  e^{ -\chi^2(D_i |\vec{\Omega},z_i,I)/2}\\
\end{equation}
where
\begin{equation}
\label{chi}
\chi^2(D_i |\vec{\Omega},z_i,I)=  \frac{\left( D_{L,obs}^i-D_{L,theo}^i(\vec{\Omega},z_i,I)\right)^2} { \sigma ^2_{D_L}}.
\end{equation}
with $D_{L,theo}^i(\vec{\Omega},z_i,I)$ denoting theoretical value of the luminosity distance corresponding to the redshift $z_i$ calculated within a cosmological model with parameters $\vec{\Omega}$.
One should note that the GW amplitude $h(t)$ measured in the detector is proportional to $D_L^{-1}$. This means that if one refers to
wave strain measurements one has to acknowledge this dependence in formulating the
likelihood. However, in our case we use the luminosity distances inferred from GW data
(along with the precision of this inference) as observables. In such case the likelihood
(2.3) is justified and indeed such kind of expression was already used by other authors in the context of GW cosmography, e.g. in \cite{Pozzo17}.

Marginalizing $P(\vec{\Omega},z_i|D_{i},I)$ over the redshift, we can write the posterior probability of cosmological parameters as:
\begin{eqnarray}
\label{Baye1}
P(\vec{\Omega}|D_{i},I)&~=~& \int_0^{z_{max}} P(\vec{\Omega},z'_i|D_{i},I)dz'_i \nonumber\\
&~=~& P(\vec{\Omega}|I)  \int_0^{z_{max}} \frac{P(D_i|\vec{\Omega},z'_i,I)P_{obs}(z'_i|\vec{\Omega},I)}{P(D_i|I)}dz'_i.
\end{eqnarray}
%
%
Given that one GW event is independent of the others, the combined posterior probability inferred from the entire set of events could be calculated.
Note that the cosmological parameter prior $P(\vec{\Omega}|I)$ is common to all events, and accordingly this prior should be used only once:
\begin{equation} \label{post}
P(\vec{\Omega}|\vec{D},I)= P(\vec{\Omega}|I) \prod_{i=1}^{n}  \int_0^{z_{max}} \frac{P(D_i|\vec{\Omega},z'_i,I)P_{obs}(z'_i|\vec{\Omega},I)}{P(D_i|I)}dz'_i.
\end{equation}

Once a set of measured luminosity distances from the GW events is obtained, the Eq.~(\ref{post}) could be calculated, provided that the prior probability distributions $P_{obs}(z_i|\vec{\Omega},I)$ and $P(\vec{\Omega}|I)$ are given.

\subsection{The prior setup}
\label{prior}
As outlined above, in order to calculate the posterior, we need to set the priors concerning cosmological
parameters $P(\vec{\Omega}|I)$ and the redshifts of GW events (i.e., $P_{obs}(z_i|\vec{\Omega},I)$).
Aiming to study the performance of cosmological inference from GW signals alone, we will set
uniform priors on $P(\vec{\Omega}|I)$) trying not to make use of values suggested by other independent
experiments.
On the other hand, the distribution of $P_{obs}(z_i|\vec{\Omega},I)$ is not straightforward and needs to be considered prudently. In this work, we derive the $P_{obs}(z_i|\vec{\Omega},I)$ by considering the intrinsic merger rate and the expected sensitivity of the ET.

In principle, the prior probability distribution of redshifts of GW sources is equivalent to the number density of the detected events as a function of redshift which have been predicted many times since the pioneering paper \cite{Finn93}. We refer the reader to more recent studies in \cite{Taylor12, ET1}. In particular, the detection rate of GWs has been calculated by \cite{ET2, ET3}, taking into account the intrinsic merger rates of the whole class of DCOs (i.e., NS-NS, BH-NS and BH-BH). These merger rates have been calculated by \cite{Dominik13} as a function of redshift using {\tt StarTrack} population synthesis evolutionary code.

The general idea of such calculation is the following. The criterion, which defines whether a DCO inspiral event is detectable, is that the value of its signal-to-noise ratio (SNR) is greater than the ET threshold (assumed as $\rho_0=$8). In general, the SNR $\rho$ for a single detector is:
\begin{equation} \label{SNR}
\rho = 8 \Theta \frac{r_0}{D_L(z_s)} \left( \frac{(1+z){\cal M}_0}{1.2 M_{\odot}} \right)^{5/6}
\sqrt {\zeta(f_{max})}
\end{equation}
where $\Theta$ is the orientation factor capturing part of sensitivity pattern due to (usually non-optimal) random relative orientation of a DCO system with respect to the detector. Four angles describe this relative orientation:
$(\theta, \phi)$ describe the direction to the binary relative to the detector, while $(\psi, \iota)$ describe the binary's orientation relative to the line-of-sight between it and the detector. The quantity $r_0$ is detector's characteristic distance parameter.  In this study, we focus on the initial ET configuration for which $r_0=1527$ Mpc.
The dimensionless function $\zeta(f_{max})$ depends only on detector's noise, its argument is the orbital frequency when the inspiral terminates, and its value is close to unity (see e.g. \cite{Taylor12}).
${\cal M}_0$ is the intrinsic chirp mass of the DCO system. Following previous work (\cite{ET1,ET2,ET3}), we assumed the chirp masses as average values for each category of DCO simulated by population synthesis:
1.2 $M_{\odot}$ for NS-NS, 3.2 $M_{\odot}$ for BH-NS and 6.7 $M_{\odot}$ for BH-BH systems. Clearly, once all other parameters are fixed, $\rho$ is a random quantity related to $\Theta$. The probability distribution for $\Theta$ calculated under the assumption of uncorrelated orientation angles $(\theta, \phi, \psi, \iota)$ has the following form:
\begin{eqnarray} \label{P_theta}
P_{\Theta}(\Theta) &=& 5 \Theta (4 - \Theta)^3 /256, \qquad {\rm if}\;\;\;
0< \Theta < 4  \\ \nonumber
P_{\Theta}(\Theta) &=& 0, \qquad {\rm otherwise} \nonumber
\end{eqnarray}

The differential inspiral rate per redshift concerning events which exceed the threshold (i.e.,  $\rho>\rho_0=$8) can be expressed as:
\begin{equation} \label{dotR}
\frac{d\dot{N}(>\rho_0)}{dz}=4\pi \left( \frac{c}{H_0} \right)^3\frac{\dot{n}_0(z_s)}{1+z_s} \; \frac{\widetilde{r}^2(z_s) }{E(z_s)} \; C_\Theta (x(z_s))
\end{equation}
where $\dot{n}_0(z_s)$ is the intrinsic coalescence rate of DCOs in the local Universe at redshift $z_s$ calculated by \cite{Dominik13} from the population synthesis code,
$C_{\Theta}(x) = \int_x^{\infty} P_{\Theta}(\Theta) d\Theta$ and $x(z, \rho) = \frac{\rho}{8} (1+z)^{1/6} \frac{c}{H_0} \frac{{\tilde r}(z)}{r_0} \left( \frac{1.2\; M_{\odot}}{{\cal M}_0} \right)^{5/6}$.
Finally, the yearly detection rate of DCO sources extending to the redshift $z_s$ can be calculated as:
\begin{equation} \label{dotN}
\dot{N} (>\rho_0|z_s)= \int_0^{z_s}  \frac{d\dot{N}(>\rho_0)}{dz}dz.
\end{equation}
Eq.~(\ref{dotN}) was used by \cite{ET2} to predict the yearly detection rate by the ET (see Table 1 and 2 therein), showing that hundreds of thousand of DCOs can be detected per year. The differential rate Eq.~(\ref{dotR}) describes the detected events distributed as a function of redshift. Therefore it could be used both to simulate the redshift distribution of the mock data and also as the prior on the redshift $P_{obs}(z_i|\vec{\Omega},I)$.

\section{Simulation and results}
\label{sim&res}
The purpose of this work is to investigate the prospects of using the future standard siren data without redshift measurements to constrain cosmological parameters. To this end, we first randomly simulate the $D_L$ data representative of what could be observed by the ET, based on the redshift distribution of these events as described in Section~\ref{prior}. Then, we apply our approach, outlined in Section~\ref{sec:theory}, to the simulated data and test its fidelity regarding the  cosmological inference.

\subsection{Mock data}
\label{simulation}
We assume flat $\Lambda$CDM Universe with $H_0 = 70$ km s$^{-1}$ Mpc$^{-1}$, $\Omega{_m} = 0.30$ as a fiducial model in our simulation. By setting this model, the redshift distribution of the mock data can be calculated using Eq.~(\ref{dotR}). We adopt the values of intrinsic inspiral rates $\dot{n}_0(z_s)$ reported by Dominik et al. \cite{Dominik13} for the whole class of DCO. Our fiducial model is the same they used.
For simplicity, we only considered the standard scenario with ``low-end" case of metallicity evolution. It has been tested (see e.g. Fig~2 in \cite{ET2}) that different choices of evolutionary scenarios would not strongly affect the final distribution.

\begin{figure*}[t!]
$\begin{array}{cc}
\includegraphics[width=0.5\textwidth]{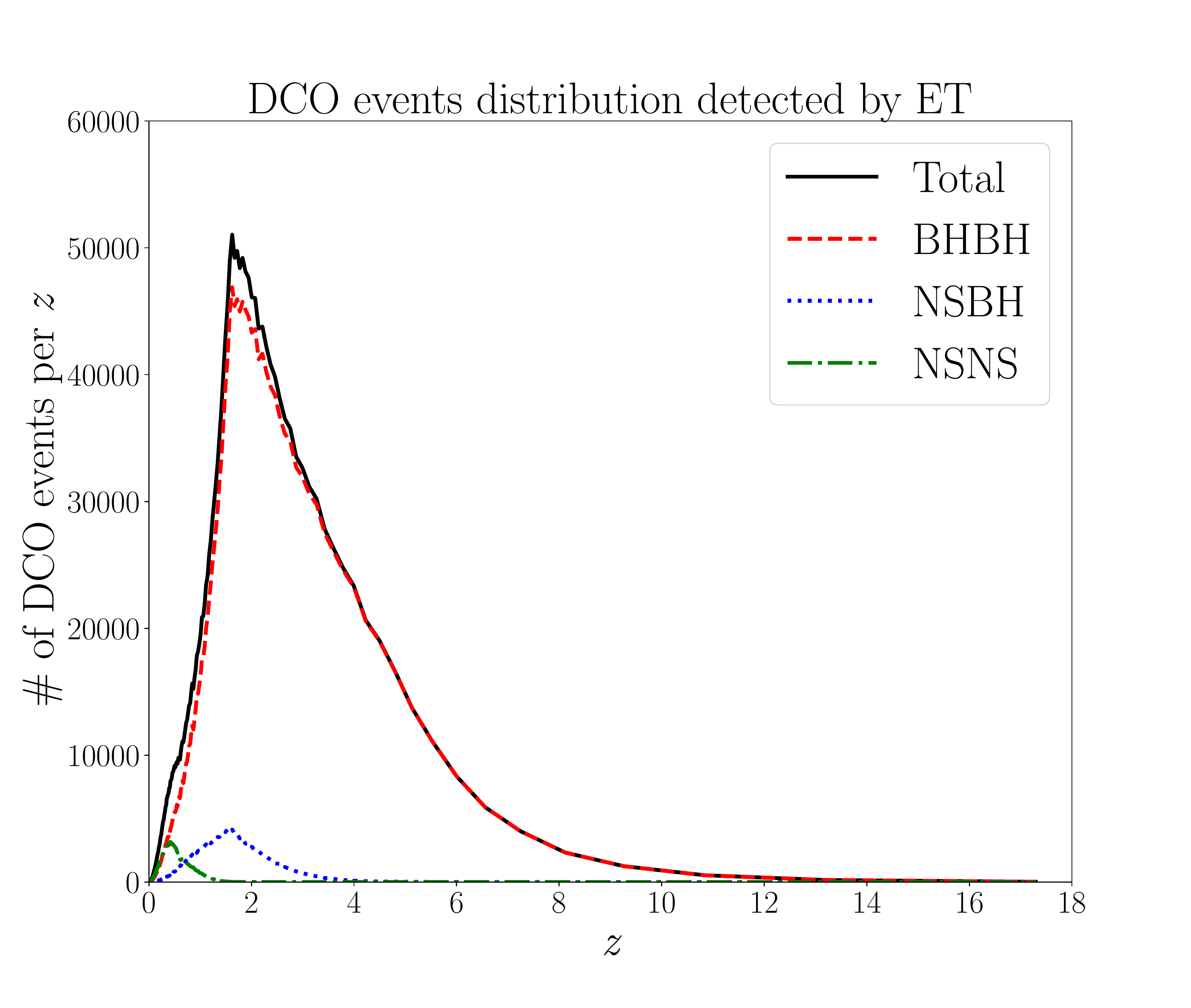}&
\includegraphics[width=0.5\textwidth]{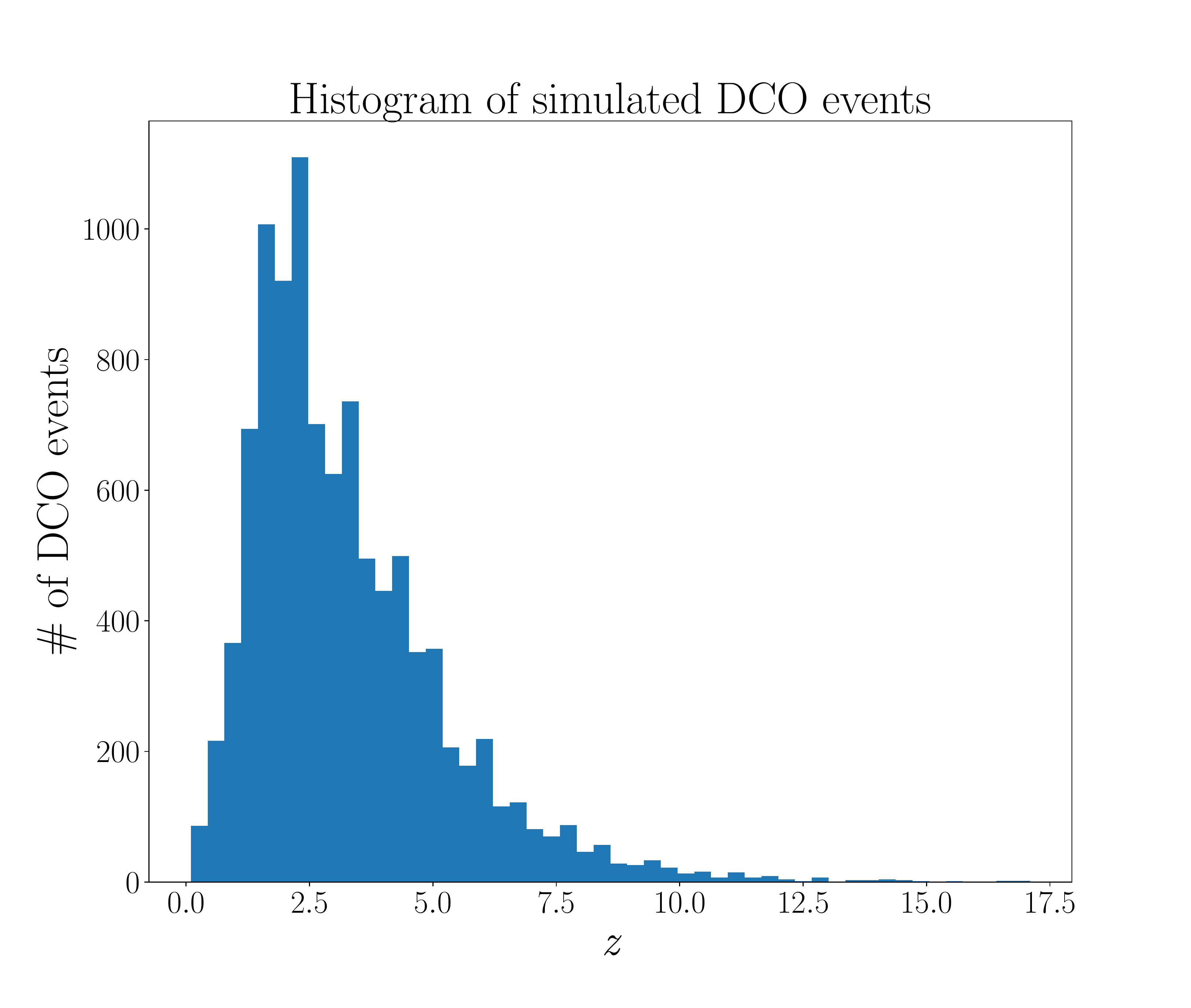}
\end{array}$
\caption[figure]{\label{fig:hist} Redshift distribution of DCO inspiral events predicted for the ET (left) and one example of the histogram of simulated DCO events (right).}
\end{figure*}

The calculated redshift distribution of DCO inspiral events predicted to be detected by the ET including NS-NS, BH-NS, BH-BH is shown in Fig.~\ref{fig:hist}.
This distribution can serve as a sampling distribution to generate the simulated redshifts of DCO systems.
Overwhelmingly, the distribution is dominated by the BH-BH systems. This is because the BH-BH systems are the predominant population of DCOs and typically have stronger signals than NS-NS, BH-NS. As an example, the histogram of redshifts obtained with 10,000 simulations is shown in Fig.~\ref{fig:hist}. This sample size is sufficient for our purpose, and at the same time, it is representative of what would be achieved very soon when the ET is put into service. In previous work \cite{ET2}, it has been estimated that the ET would register about $10^4-10^5$ inspiral DCO events per year.

Following the common practice, mock luminosity distance is generated as $D_{L,sim}(z_s)=D_{L,fid}(z_s)+\mathrm{N}(0,\sigma)$, where $D_{L,sim}(z_s)$ and $D_{L,fid}(z_s)$ are  simulated and fiducial values of the luminosity distance at a given redshift, respectively. Assuming the fiducial cosmological model as a `true' one, the values of luminosity distance $(D_{L,fid}(z))$ at the corresponding redshift can be calculated within such model. The $\mathrm{N}(0,\sigma)$ term is the Gaussian random variable with zero mean and variance $\sigma$ corresponding to the uncertainty regarding the luminosity distance measurement. Given the values of uncertainty level, one can randomly generate the simulated values of $D_{L,sim}$.

\subsection{Cosmological inference}
\label{cos_infer}

In this section, we investigate posterior distributions of cosmological parameters from the
analysis of simulated mock data using our approach based on Eq.~(\ref{Baye1}) and~(\ref{post}).
We consider two simplest cosmological models with the following expansion rates:
\begin{eqnarray}
\label{cosm_model}
&~H(z)~&=H_0\sqrt{ \Omega_m(1+z)^3+(1-\Omega_m)}, \\
&~H(z)~&=H_0 \sqrt{ \Omega_m(1+z)^3+(1-\Omega_m)(1+z)^{3(1+w)}},
\end{eqnarray}
which phenomenologically describe the dark energy modeled as a perfect fluid with barotropic equation of state:
$p = -\rho$ and $p = w\rho$, 
respectively\footnote{These models are known as flat $\Lambda$CDM and $w$CDM, respectively.}.

We performed the Monte Carlo probability maximization by first repeatedly simulating a set of mock data realizations with random noise and of a sufficiently big size; each realization contained 10,000 events. We then derived the maximization distribution of the posterior for the parameters $\vec{\Omega}$ using Eq.~(\ref{post}) based on the realizations. The simulation process continued until the maximization distribution was stable. In this study, only the luminosity distance is considered as the observed data, and the uncertainty of this distance would not usually be perfectly known. Thus, in our analysis, the uncertainty is assumed as a parameter which should be marginalized over in the final result. Moreover, these uncertainty levels would affect the posterior; thus we adopt three different uncertainty levels randomly distributed as $U([5\%, 10\%])$, $U([5\%, 15\%])$, $U([5\%, 20\%])$, respectively. We infer their corresponding cosmological inference, in sequence.

The inference for two cosmological models mentioned above is given in the following subsections.

\subsubsection{The $\Lambda$CDM model}
\label{lcdm}
We assume uniform priors: $H_0 \sim U([45, 95])$ and $\Omega_m \sim U([0.1, 0.55])$ for the cosmological parameters. Since the uncertainty level of the $D_L$ is not perfectly unknown, we adopt the uncertainty level as $U([5\%, 20\%])$ as the universal prior.

\begin{figure*}
\centering
\begin{tabular}{ccc}
\subfloat[$D_L$ uncertainty distributed as $U(5\%, 10\%)$.]{\includegraphics[width=0.3\textwidth]{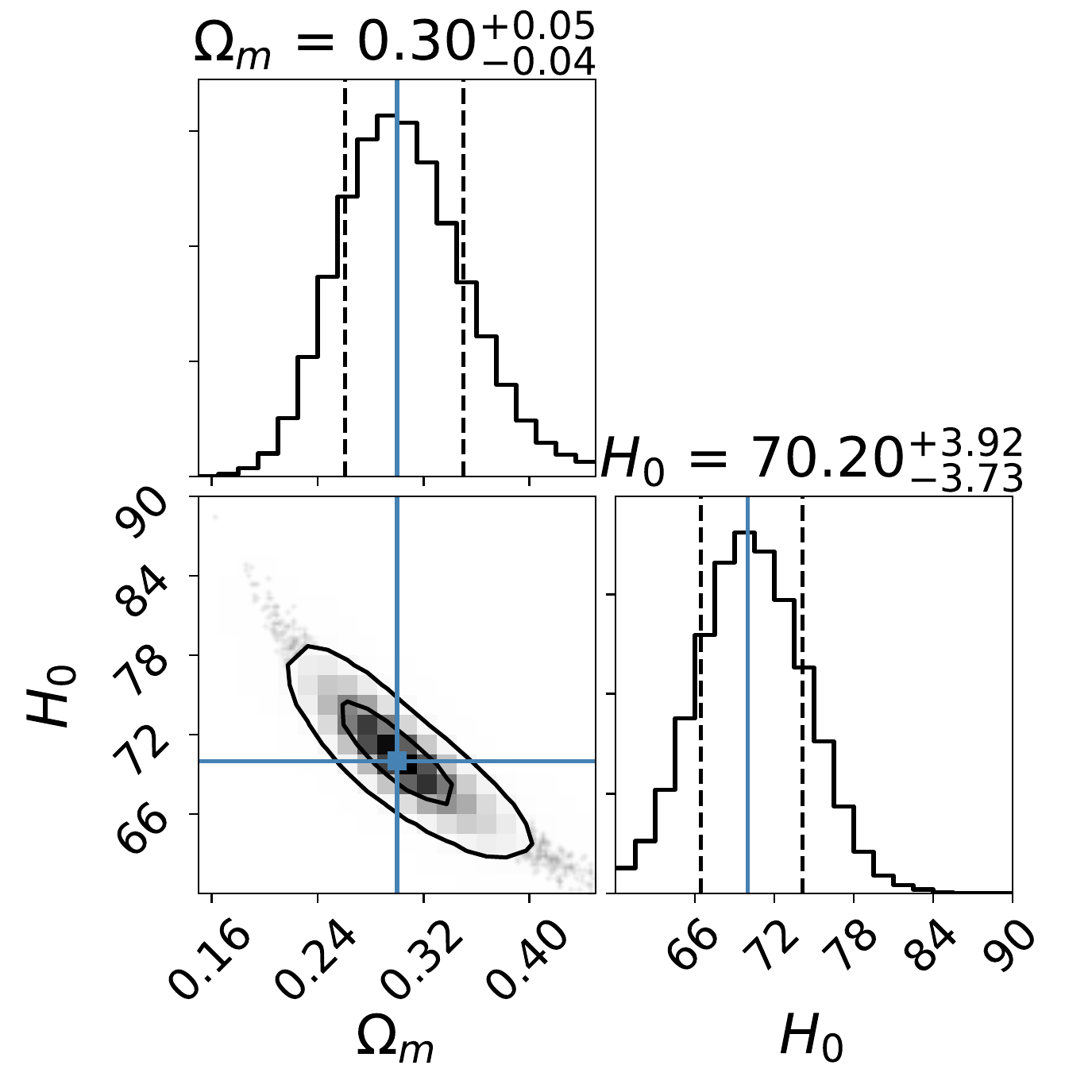}}&
\subfloat[$D_L$ uncertainty distributed as $U(5\%, 15\%)$.]{\includegraphics[width=0.3\textwidth]{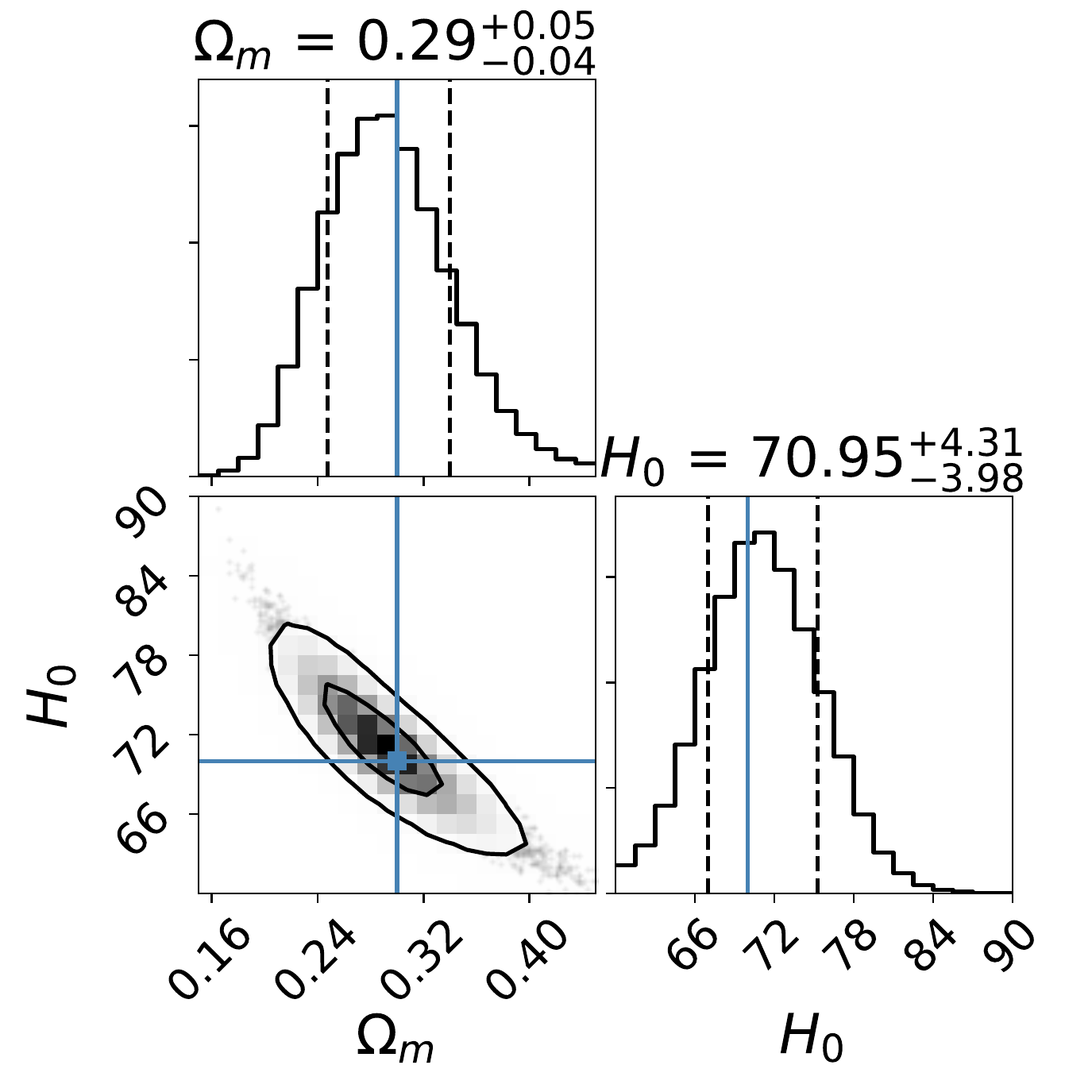}}&
\subfloat[$D_L$ uncertainty distributed as $U(5\%, 20\%)$.]{\includegraphics[width=0.3\textwidth]{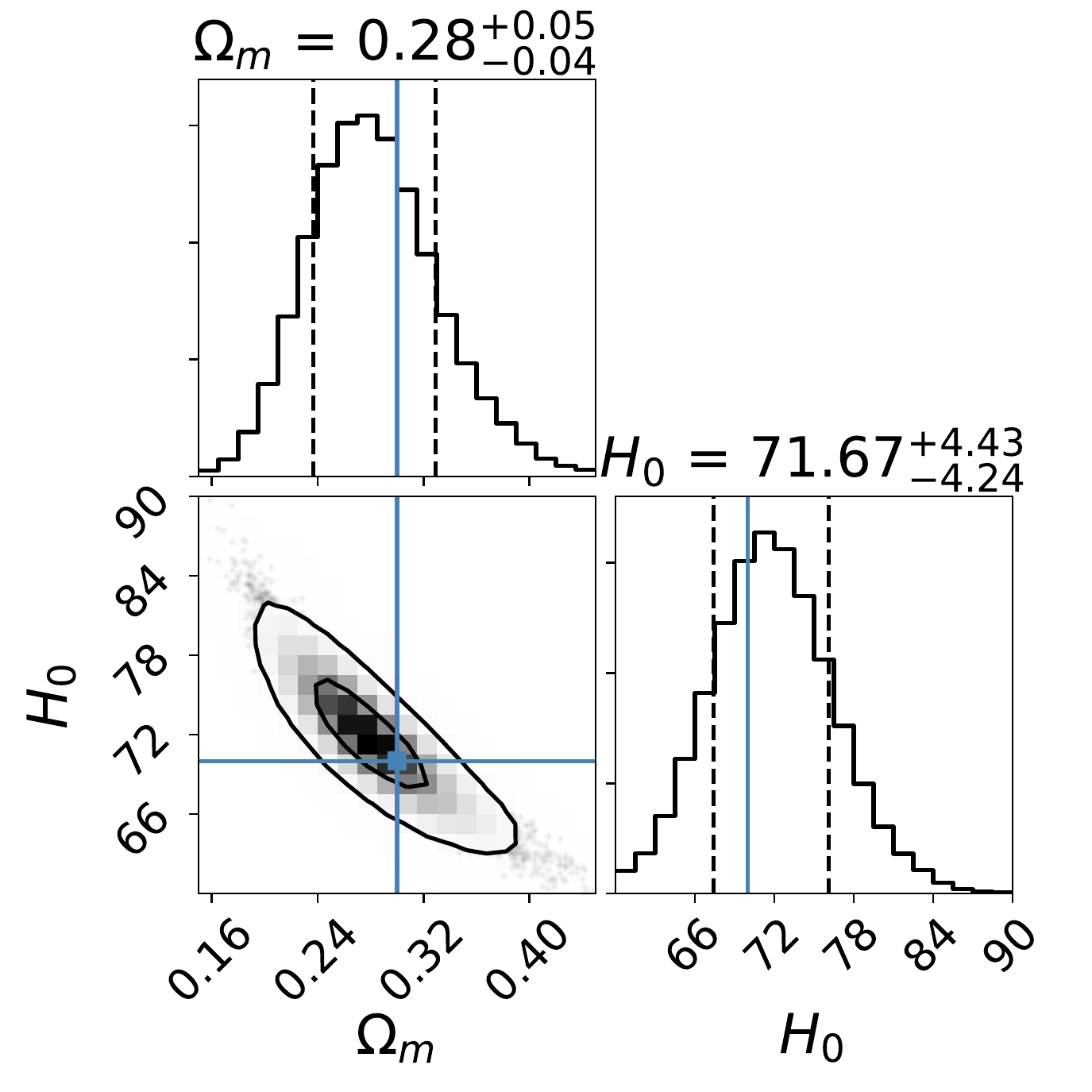}}\\
\end{tabular}
\caption{\label{fig:lcdm_infer} Illustration of the maximization distribution of the posterior for the cosmological parameters in the flat $\Lambda$CDM model
based on $\sim5,000$ realizations of datasets, each dataset contain 10,000 events.
In the simulation, $D_L$ are set with different uncertainty levels. In the fitting, the universal prior of uncertainty level is adopted as U([5\%,20\%]).
 The contour regions denote the $68\%$ (1-$\sigma$) and $95\%$ (2-$\sigma$) confidence region. The blue lines denote the true values of the fiducial model parameters.}
\end{figure*}

The confidence contours and marginalized Probability Distribution Functions (PDFs) are shown in Fig.~\ref{fig:lcdm_infer}-(a). In order to study the precision and accuracy of cosmological inference as a function of data quality, we set the uncertainty of $D_L$ at different levels as mentioned above.
We see that the reliable inference (within $68\%$ (1-$\sigma$) confidence) could be achieved at each uncertainty level. Not surprisingly, with the higher uncertainty level, the scatter of $H_0$ realizations is slightly larger with the central value slightly shifted from the true point. We tested that adopting the prior uncertainty levels other than $U([5\%, 20\%])$ could shift the central region toward different directions. Yet, all these modifications resulted with an inference in agreement with true values within 1-$\sigma$.

Besides the uncertainty level, the confidence regions for the inferred cosmological parameters are related to the number of detected events. As we discussed in Section~\ref{into}, the precision of $H_0$ measurement is necessary to shed light on the tension between \textit{Planck} and local probes. Therefore we also investigated how many data on $D_L$ would be required to achieve a percent precision for the Hubble constant.
Fixing the uncertainty level for $D_L$ at $10\%$, we increased the size of the
data gradually from $2\times10^3$ to $1\times10^5$ and obtained the corresponding 1-$\sigma$ confidence region of inferred $H_0$, as listed in Tab.~\ref{pmH0}. The result shows that the inference of $H_0$ with $\sim1\%$ precision 
 requires $1\times10^5$ samples of $D_L$.

\subsubsection{The $w$CDM model}

In the $w$CDM model, the equation of state coefficient $w$ is a free parameter. Therefore, besides $H_0$ and $\Omega_m$, for which we assume the same uniform priors as in $\Lambda$CDM case, we should set a prior on $w$ as well. To calculate the posterior, we assumed a uniform prior $w \sim U([-2.0, -0.5])$.

\begin{figure*}
\centering
\begin{tabular}{ccc}
\subfloat[$D_L$ uncertainty distributed as $U(5\%, 10\%)$.]{\includegraphics[width=0.3\textwidth]{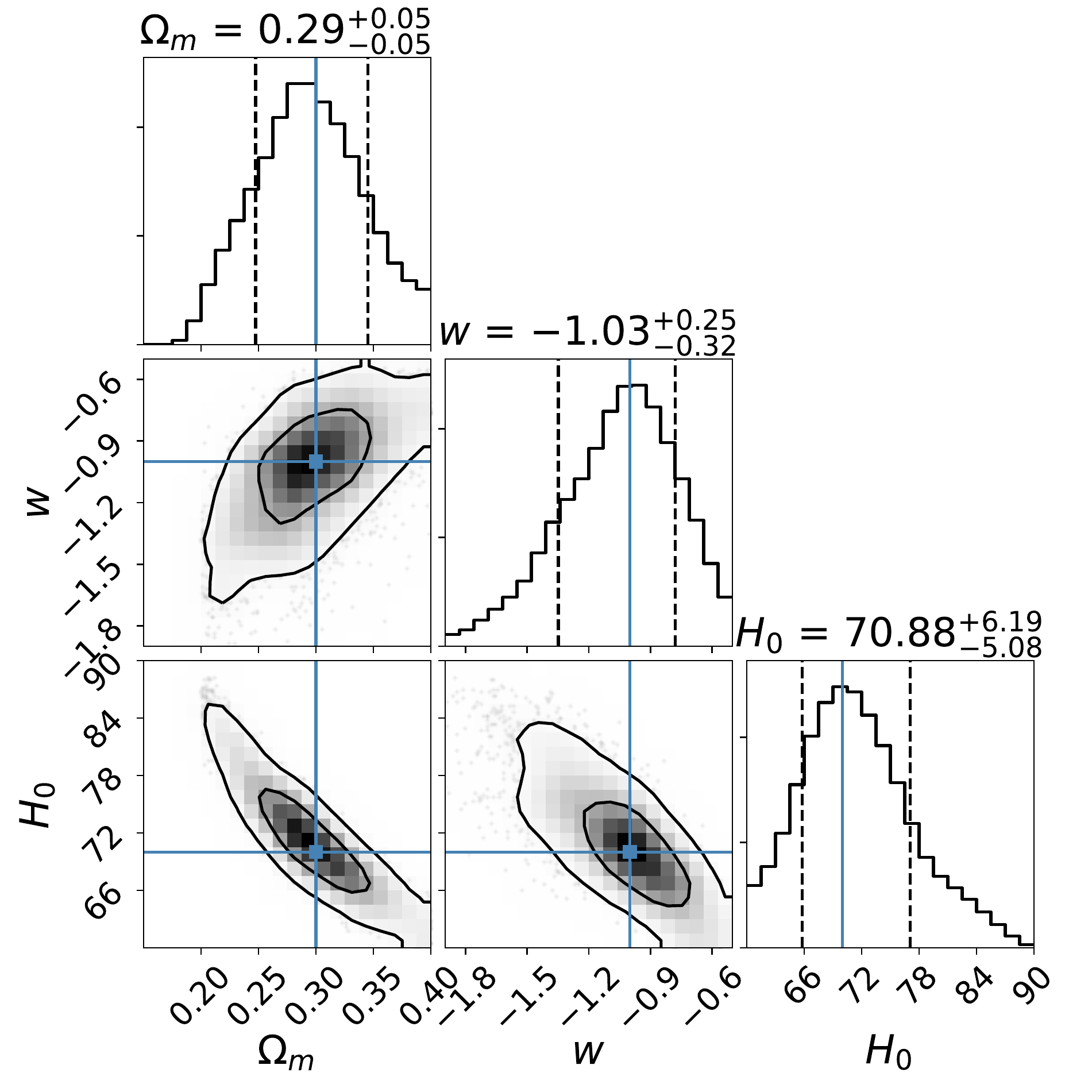}}&
\subfloat[$D_L$ uncertainty distributed as $U(5\%, 15\%)$.]{\includegraphics[width=0.3\textwidth]{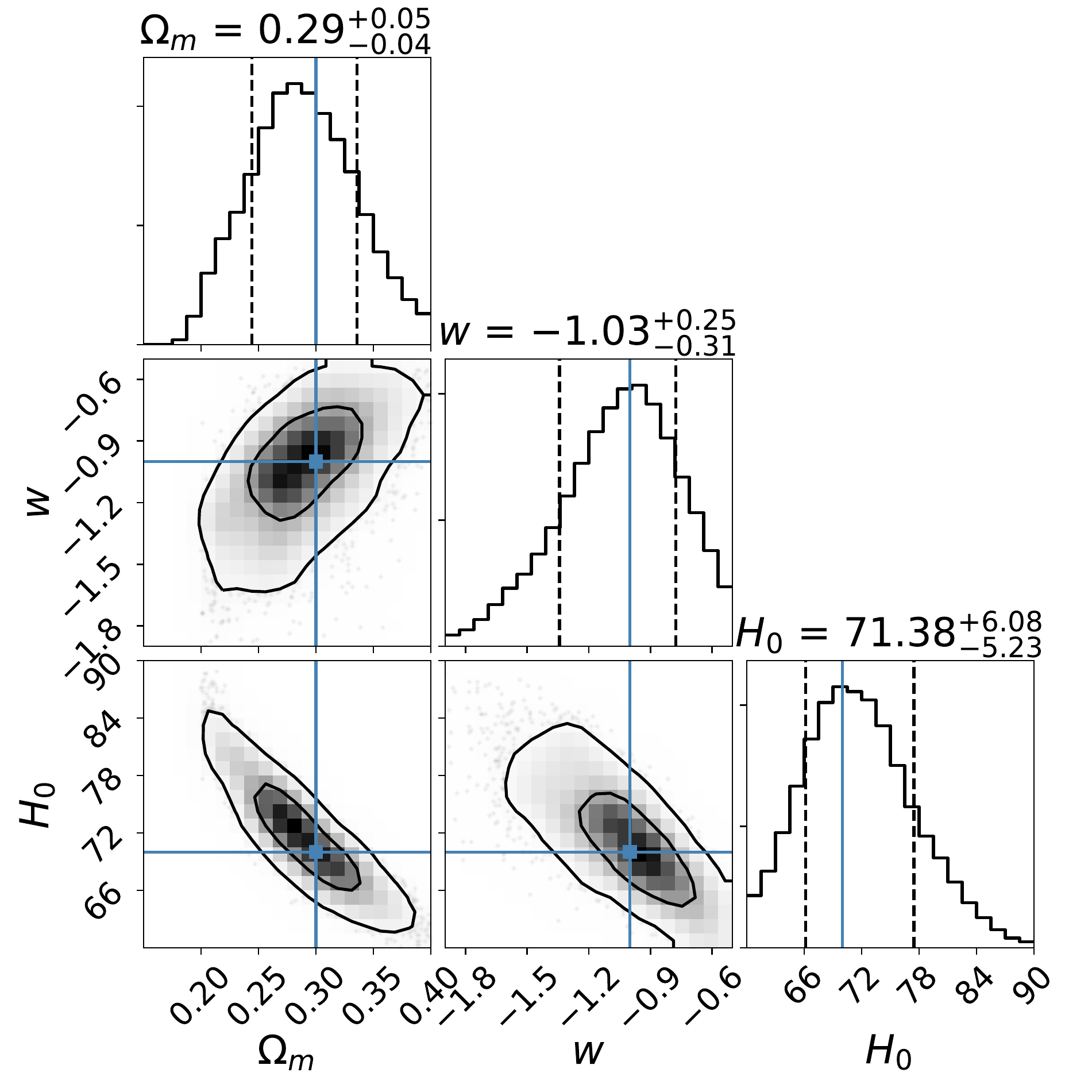}}&
\subfloat[$D_L$ uncertainty distributed as $U(5\%, 15\%)$.]{\includegraphics[width=0.3\textwidth]{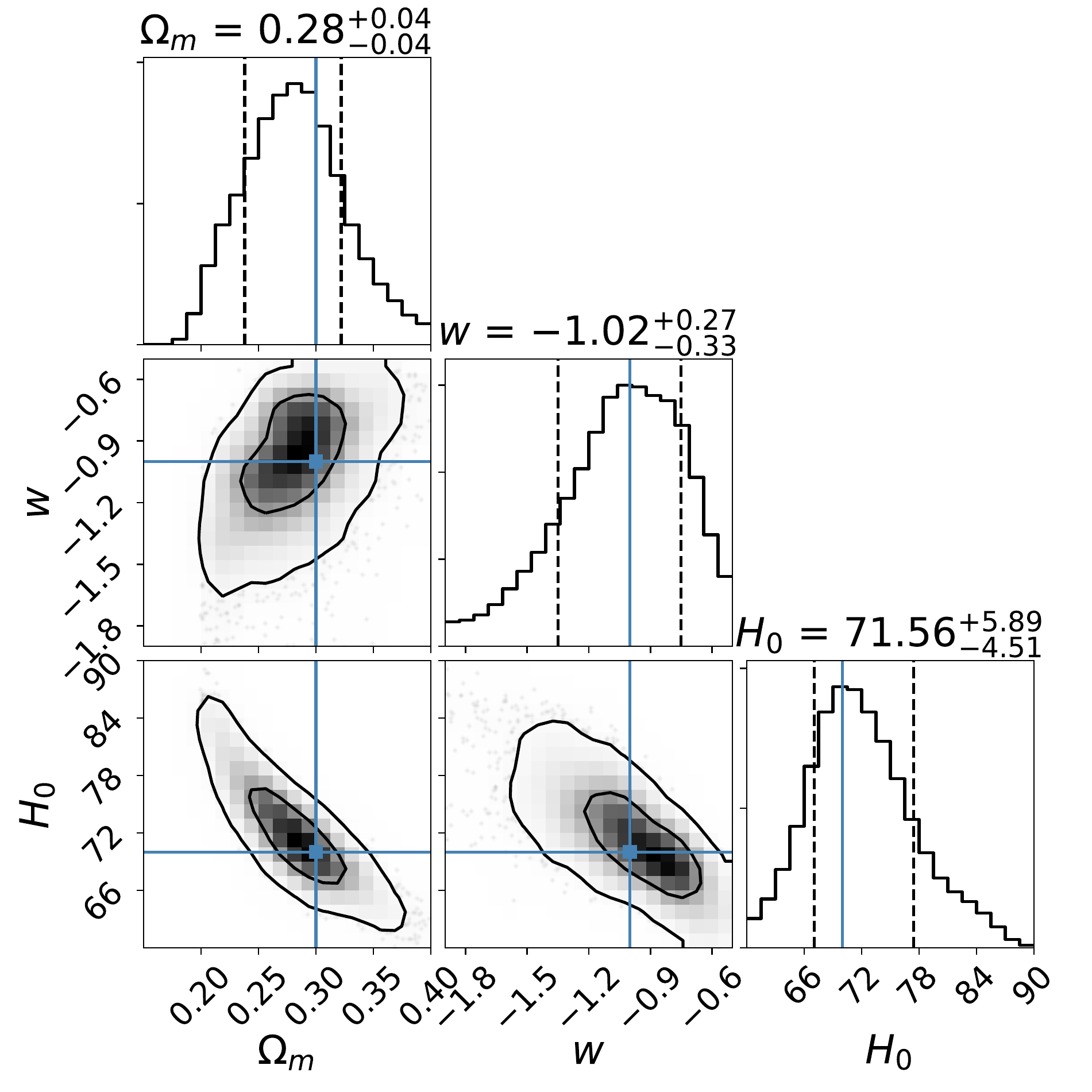}}\\
\end{tabular}
\caption{\label{fig:wcdm_infer} Results for $w$CDM model. 
}
\end{figure*}

Since there is a strong degeneracy between the equation of state for dark energy ($w$ - parameter) and other parameters, the posterior distributions are supposed to be wider when $w$ is set free. Indeed, in Fig.~\ref{fig:wcdm_infer} we present the results
  for the $w$CDM model with wider confidence contours. Despite of this degeneracy, the results indicate that our approach is still able to recover cosmological  parameters within 1-$\sigma$ for the distance uncertainty level up to 20\%.

Concerning the uncertainty of cosmological inference as a function of sample size (Tab.~\ref{pmH0}), one can see that with the biggest sample of $100\times10^3$ measurements, the $1-\sigma$ confidence region for $H_0$  would be as big as $\pm5.2$, which corresponds to $\sim 7\%$ precision.

\begin{table}
\centering
    \caption{The 1-$\sigma$ confidence region for $H_0$ as a function of sample size. The $D_L$ uncertainty is distributed as $U(5\%, 15\%)$.}\label{pmH0}
     \begin{tabular}{cccccc}
     \hline\hline
     & \multicolumn{5}{c}{$\Delta H_0$ }\\  \hline
      N ($\times 10^3$) & 2 & 5 & 10 & 50 &  100\\    \hline\hline
     $\Lambda$CDM & $\pm6.5$ & $\pm4.3$ & $\pm3.4$ & $\pm1.4$ & $\pm 1.0$ \\
     $w$CDM & $\pm10.6$ & $\pm10.3$ & $\pm9.5$ & $\pm6.1$ & $\pm5.2$\\
     \hline
     \end{tabular}
\end{table}

\section{Discussion and conclusion}
\label{sec:conclusion}
In this paper, we investigated the prospects of using gravitational waves from inspiralling compact binaries as standard sirens for the cosmological inference. Though the redshift $z$ of such GW events would be unknown -- unmeasurable from the GW waveforms, and very hard to obtain by EM counterpart identification for sources at greater distances and for BH systems --
we proved that this inference could be achieved as long as the sources redshift probability distribution could be provided.

Using the Bayes theorem, we constructed the posterior of cosmological parameters using the redshift distribution as a  prior Eq.~(\ref{post}). We have shown that this prior, i.e. $P_{obs}(z_i|\vec{\Omega},I)$, is predictable, given the intrinsic merger rate of DCO events and the expected sensitivity of the detector, i.e. the ET in the case we discussed. Then, we estimated the precision and accuracy of our approach using simulated mock data generated by combining the $P_{obs}(z_i|\vec{\Omega},I)$ with the fiducial cosmological model.

We repeatedly generated the realizations of datasets each containing 10,000 systems and computed the maximization distribution for the parameters ($\vec{\Omega}$) of $\Lambda$CDM and $w$CDM model. Because the data were simulated from the fiducial model, the true values of cosmological parameters were assumed as known. Therefore the inferred values of these parameters allowed to study both the precision and accuracy of the inference as well as their changes as a function of data quality, i.e. the uncertainty of $D_L$ measurements. We stress again that in this work, we assumed that the only observable quantity was the luminosity distance whose uncertainty level was not perfectly known. Consequently, the luminosity distance uncertainty was assumed as a free parameter in the analysis and marginalized over. Our results show that one can obtain the non-biased cosmological inference at different $D_L$ uncertainty levels up to 20\%
in agreement with pre-assumed true values within 1-$\sigma$ level.

We also investigated the confidence regions for the inferred cosmological parameters as a function of sample size. We
found that if one aims at the $H_0$ measurement contributing to the resolution of the tension in $\Lambda$CDM model between Planck and other low redshift measurements, one needs a sample size of $\sim1\times10^5$ events (see. Tab.~\ref{pmH0}). Even though it seems large, such a sample size could be provided in one year of successful operation of the ET.

In the literature concerning LIGO/Virgo or LISA detectors, different concepts concerning redshift priors have been discussed. Mostly, the idea there was to incorporate all potential host galaxies \cite{Pozzo2012, Pozzo17, Chen17} or clusters \cite{MacLeod2008} from wide-field sky surveys such as the SDSS.
Recently, \cite{Fishbach2018} performed a statistical standard siren analysis of GW170817 which did not utilize knowledge of NGC 4993 as the unique host galaxy. By weighting the host galaxies by stellar mass or star-formation rate they obtained consistent results with potentially tighter constraints.
Admittedly, such statistical methods were claimed to be able to constrain $H_0$ to several percent levels. Such approaches are only applicable
whenever the redshift proxies can be well assessed or the electromagnetic counterpart is detected. However, the GW events registered in the era of the ET would come from very high redshifts at which host galaxies would not be available to the wide-field surveys. Hence, it is possible that the approach proposed in this paper might be the only option in the era of 3rd generation of GW detectors.

\begin{figure*}
\centering
\includegraphics[width=0.7\textwidth]{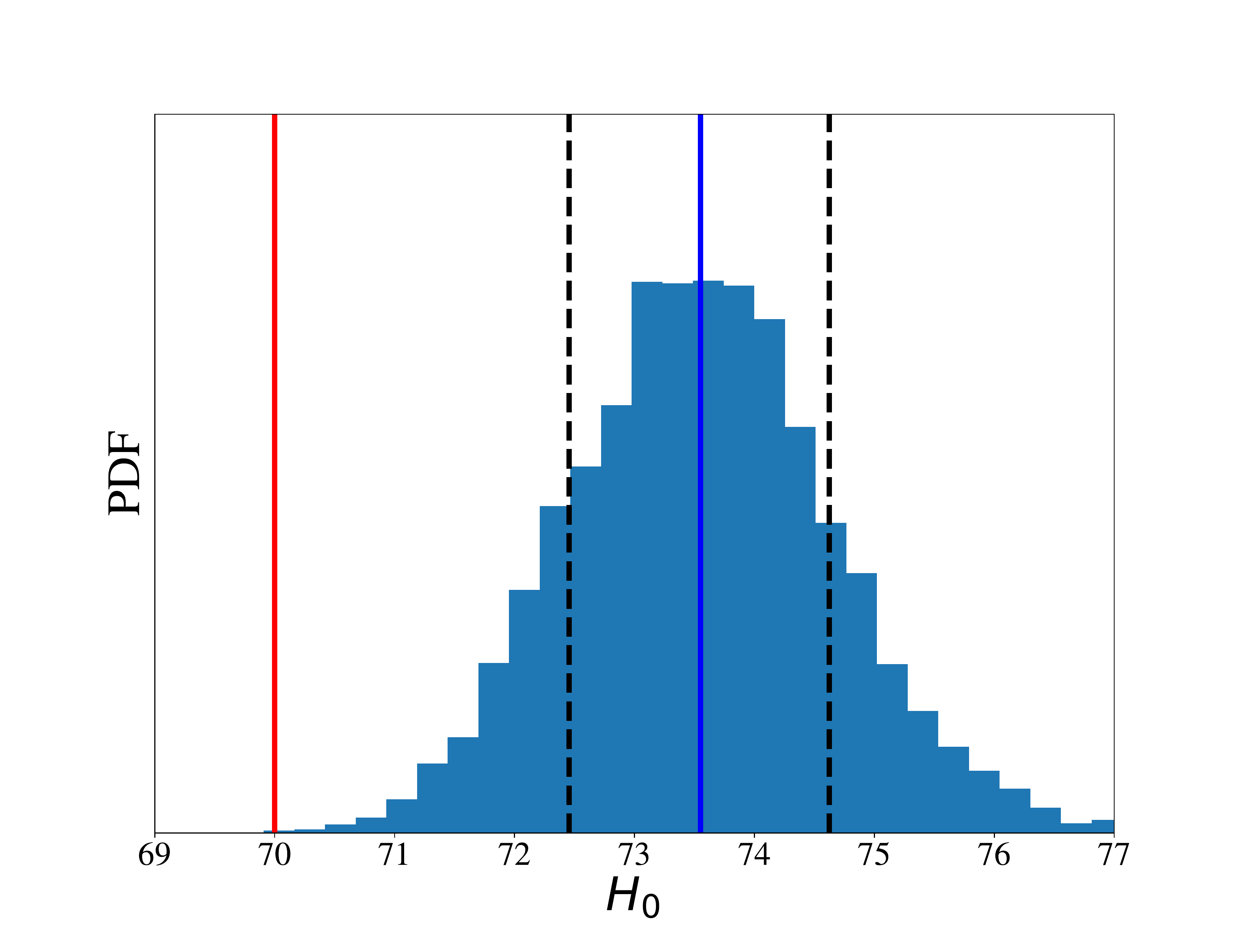}
\caption{\label{fig:1D-pdf} The inferred $H_0$ when the standard scenario for $\dot{n}_0(z_s)$ was used to generate the mock data, but the delayed SN scenario was used as a prior. The red line is the true value of $H_0$ used in mock data generation.
}
\end{figure*}

In this work, we used only one particular population synthesis model of inspiral rates $\dot{n}_0(z_s)$ (i.e. the ``low-metallicity'' standard scenario) to generate the mock data, and then to derive cosmological parameters. Even though the difference between each population model by Dominik et al. \cite{Dominik13} is small, it can be expected that this small difference could be amplified by the selection effects inherent to the GW observations.
This means that if the wrong prior of $\dot{n}_0(z_s)$ is assumed, an extra bias would be introduced. To test this, we took for the simulations the $\dot{n}_0(z_s)$ according to the standard scenario, but used the delayed SN scenario as a prior for the inference.
In order to directly observe the bias on $H_0$ induced by such mismatch in assumptions, we fixed the value of $\Omega_m$ during the fitting. The result shown in Fig.~\ref{fig:1D-pdf} from which the bias on the inferred $H_0$ is clearly seen.
In the future, we hope that $\dot{n}_0(z_s)$ would be known better, following better understanding of the DCO evolution and refinement of population synthesis models based on existing and forthcoming GW detections by LIGO/Virgo.
Let us also remark that our approach could also be applied to constraining the right scenario for $\dot{n}_0(z_s)$ --- one can set the cosmology as prior and select the best scenario. Assessment of this idea would require more extensive tests and simulations, and is left for the future work.

Furthermore, the $\dot{n}_0(z_s)$ also depends on the cosmological model, since the population synthesis models predict coalescence rate as a function of time, and time-redshift relation should be used. We have avoided this problem by adopting the same cosmology as Dominik et al. \cite{Dominik13} have used.
This issue deserves more comprehensive studies in the future using parametrized models for the DCO population.

Even though it was shown in \cite{kai2017} that with about 10 lensed GW and EM signals (realistic number for the ET) one would achieve a sub-percent accuracy of the $H_0$ measurement, yet it would not be fast and easy to gather such a sample. In particular, the pipelines to identify lensed GW events are still under development.
On the contrary, luminosity distances inferred from the inspiral waveforms would be routinely measured in significant numbers quickly building up the samples we discussed. Therefore it would be promising to develop further the method we proposed.

\acknowledgments
We would like to express our deep gratitude to the referee for a thorough reading of our paper at each stage of revisions and his/her time devoted to constructively discuss the issues that needed improvements. Especially the comments regarding the bias associated with assumptions of uncertainty level and population synthesis scenario are gratefully acknowledged. This constructive engagement of the referee allowed to improve the paper substantially.
We thank Xi-Long Fan for contributing to the formulation of this work; unfortunately, he did not wish to be
an author because of restrictions required by the LIGO Scientific Collaboration policies.
This work was supported by the National Basic Science Program (Project 973) of China under (Grant No. 2014CB845800), the National Natural Science Foundation of China under Grants Nos. 11633001 and 11373014, the Strategic Priority Research Program of the Chinese Academy of Sciences, Grant No. XDB23000000 and the Interdiscipline Research Funds of Beijing Normal University.
X. Ding acknowledges support by China Postdoctoral Science Foundation Funded Project (No. 2017M622501).
M.B. was supported by the Key Foreign Expert Program for the Central Universities No. X2018002
K. Liao was supported by the National Natural Science Foundation of China (NSFC) No. 11603015.
Z. Li was supported by NSFC under Grants Nos. 11505008.

\providecommand{\href}[2]{#2}\begingroup\raggedright\endgroup

\end{document}